\documentclass[12pt]{iopart}
\usepackage{graphicx}

\begin{document}
\title{Cosmology with variable parameters and effective equation of state for Dark Energy}
\author{Joan Sol\`{a}}
\address{HEP Group, Departament d'Estructura i Constituents de la Mat\`{e}ria, Universitat de Barcelona,
Av. Diagonal 647,  08028 Barcelona, Catalonia, Spain \\
and C.E.R. for Astrophysics, Particle Physics and Cosmology
\footnote{Associated with Instituto de Ciencias del Espacio -
CSIC} } \ead{sola@ifae.es}
\author{Hrvoje \v{S}tefan\v{c}i\'{c} \footnote{On leave of absence from the Theoretical Physics Division,
Rudjer Bo\v{s}kovi\'{c} Institute, Zagreb, Croatia}}
\address{HEP Group, Departament d'Estructura i Constituents de la Mat\`{e}ria, Universitat de Barcelona,
Av. Diagonal 647,  08028 Barcelona, Catalonia, Spain}
\ead{stefancic@ecm.ub.es}
\begin{abstract}
A cosmological constant, $\Lambda$, is the most natural candidate
to explain the origin of the dark energy (DE) component in the
Universe. However, due to experimental evidence that the equation
of state (EOS) of the DE could be evolving with time/redshift
(including the possibility that it might behave phantom-like near
our time) has led theorists to emphasize that there might be a
dynamical field (or some suitable combination of them) that could
explain the behavior of the DE. While this is of course one
possibility, here we show that there is no imperative need to
invoke such dynamical fields and that a variable cosmological
constant (including perhaps a variable Newton's constant too) may
account in a natural way for all these features.
\end{abstract}

\section{Introduction}
The phenomenon of the accelerated expansion of the universe is
presently one of the central issues of both observational and
theoretical cosmology. A number of diverse cosmological
observations\,\cite{Supernovae,WMAP03LSS} have by now established
the accelerated nature of the present expansion of the universe
and even provided additional information on the
deceleration/acceleration transition and the redshift dependence
of the expansion of the universe. From the theoretical side, the
sole fact that the universe is presently accelerating, and may
continue to do so, has triggered many studies. Some particularly
interesting possibilities include  braneworld models of the
late-time cosmic acceleration \cite{braneworld}. The real
theoretical challenge, however, lies in understanding the dynamics
leading to the accelerated expansion of the universe. Despite the
fact that many promising models have been proposed, the
fundamental nature of the accelerating mechanism remains presently
unknown. The attempts towards shedding some light on the cause of
the acceleration of the universe employ a broad range of concepts
in many theoretical frameworks. The most widely used and the
conceptually simplest option assumes the existence of {\em dark
energy} (DE), the cosmic component with the negative pressure.
Dark energy is a very useful concept since it encodes all our
ignorance on the acceleration of the universe in a single cosmic
component. Furthermore, DE can also be used as an effective
description of other mechanisms of the acceleration of the
universe\,\cite{Eqos}. A traditional candidate for the role of DE
is the cosmological constant (CC), $\Lambda$. In fact, the
cosmological FRW model with cold dark matter and $\Lambda$ as the
DE component (the so-called $\Lambda$CDM cosmology) fits the data
reasonably well\,\cite{Supernovae,WMAP03LSS}. Theoretically,
however, the CC as a DE candidate faces a formidable problem
related to the very many ($55$ at least) orders of magnitude
difference of its predicted value in quantum field theory (QFT)
and the observed value\,\cite{CCP}. This huge discrepancy leads
indeed to the cosmological challenge of the millennium and calls
for a more profound treatment of the CC problem\,\cite{Paddy}.
The many inconsistencies related to the ``$\Lambda$ conundrum''
have led to the development of the dynamical DE models. The tacit
assumption of all these models is that $\Lambda$ does not
contribute to the DE or that it vanishes, which merely
circumvents the CC problem. The dynamical DE models, however,
incorporate the advantage that they approach the modeling of the
mysterious dark energy component in a more general way, allowing
its properties to vary with the expansion. Notwithstanding this
welcome feature, the necessity of a (severe) fine-tuning of the
parameters of these models in order that the measured value of
the DE coincides with the predicted value does persist and it is
in no way less worrisome than in the CC case. The list of
dynamical DE models comprise, among others, {\em quintessence}
\cite{Quintessence}, {\em phantom energy} \cite{phantom},{\em
Chaplygin gas}\,\cite{Chaplygin} etc\,\cite{CCP,Paddy}. Very often
these models of DE are realized in terms of dynamical scalar
field(s). These fields were actually introduced on more or less
phenomenological grounds to deal with the CC problem long ago
\,\cite{Dolgov,PSW}. In particular, in the popular quintessence
approach the scalar fields and their parameters/scales are
totally unrelated to the known particle physics fields, and as a
consequence an obvious connection to fundamental physics is
lacking. Not only so, in all these models the scalar field
potentials just concocted to describe an acceptable form of DE
are non-renormalizable by power counting and, what is worse, they
are essentially field-theoretically unmotivated.

In this paper we take the point of view that these DE models are
not more fundamental than the original $\Lambda$ model. However,
we exploit the most useful property of the former, namely the
dynamical character of the DE. Therefore, we generalize the CC
concept allowing the variability of the $\Lambda$ term (and
possibly also of the gravitational coupling $G$) with the cosmic
time, both assumptions being perfectly compatible with the
cosmological principle and can be realized, as we shall see, in a
fully covariant way. The support for such a generalization of the
$\Lambda$ term comes from QFT on curved space-time
\cite{JHEPCC1,Babic} and/or quantum gravity approaches
\cite{Reuter}. Here, however, we do not derive the variability
of  $\Lambda$ and $G$ from these models, but discuss the general
implications of this variability. Our conclusions will therefore
be valid for any specific model of the mentioned type. Let us
emphasize that this approach embodies many virtues: i) the
problem of the CC is dealt with instead of circumventing it; ii)
the variability of $\Lambda$ may shed some light on the observed
value of $\Lambda$; iii) a variable $\Lambda$ is indeed a type of
dynamical DE, and will be handled here in the language of the
effective DE picture. We will show in a precise way how a
cosmology with variable parameters $(\Lambda,G)$ can mimic a
dynamical DE model and produce a non-trivial {\em effective}
equation of state (EOS), $p_D=w_{\rm eff}\,\rho_D$, beyond the
naively expected one for the $\Lambda$ term ($w_{\Lambda}=-1$).
Since the EOS parametrization is widely used in all planned
experiments aiming at a precise study of the DE\,\cite{Eqos}, such
correspondence can be useful and it may even unveil some
unexpected limitations of the EOS method for characterizing the
ultimate nature of the DE.

\section{Dark energy picture versus variable cosmological parameters picture}
Before presenting the general procedure for obtaining the
effective dark energy EOS corresponding to a generic variable
$\Lambda$ and $G$ model, we briefly discuss the frameworks of two
approaches which we call for short the {\em DE picture} and the
{\em variable CC picture}. The {\em DE picture} assumes the
existence of two separately conserved cosmological energy density
components, the matter-radiation component and the DE component.
In this picture the Einstein field equations
\begin{equation}
\label{eq:Ein1} R_{\mu \nu} -\frac{1}{2} g_{\mu \nu} R = 8 \pi G\,
\tilde{T}_{\mu \nu}
\end{equation}
have a total energy-momentum tensor of the form $\tilde{T}_{\mu
\nu} = T^{s}_{\mu \nu} + T^{D}_{\mu \nu}$. The standard
energy-momentum tensors of matter-radiation  $T^{s}_{\mu \nu}$
and the one for dark energy $T^{D}_{\mu \nu}$ are conserved
separately. In the framework of FRW metric, the conservation of
the energy-momentum tensor for matter-radiation, $\nabla^{\mu}
T^{s}_{\mu \nu} = 0$, leads to the standard conservation law
\begin{equation}
\label{eq:conservNR} \frac{d \rho_{s}}{dt} + \alpha H_{D}\,
\rho_{s} = 0\, .
\end{equation}
Here $\alpha = 3 (1+\omega_{s})$ where $\omega_{s}=0$ and $1/3$
for nonrelativistic matter and radiation, respectively. In the
standard {\em DE picture} the assumption that $\nabla^{\mu}
T^{D}_{\mu \nu} = 0$ results in the additional conservation law
for the DE:
\begin{equation}
\label{eq:conservDE} \frac{d \rho_{D}}{dt} + 3 (1 + w_{\rm eff})
H_{D}\, \rho_{D} = 0\, .
\end{equation}
Generally the parameter $w_{\rm eff}$ is redshift dependent,
$w_{\rm eff}=w_{\rm eff}(z)$. The Hubble parameter in this
picture is defined by the Friedmann equation (assuming zero
spatial curvature)
\begin{equation}\label{DEpicture}
H_D^2=\frac{8\pi G_0}{3}(\rho_s+\rho_D) \, ,
\end{equation}
where a subscript $D$ has been appended to $H$ to distinguish the
expression (\ref{DEpicture}) from its counterpart in the variable
CC picture -- see (\ref{CCpicture}) below. The solution of the
conservation laws (\ref{eq:conservNR}) and (\ref{eq:conservDE})
results in the following scaling laws for the components of the
model:
\begin{equation}
\label{eq:rhos} \rho_s(z)=\rho_s(0)\left(1+z\right)^{\alpha}
\end{equation}
and
\begin{eqnarray}\label{zeta}
&&\rho_D(z)=\rho_D(0)\,\zeta(z)\,,\\
&&\zeta(z)\equiv\,\exp\left\{3\,\int_0^z\,dz' \frac{1+w_{\rm
eff}(z')}{1+z'}\right\}\nonumber\,.
\end{eqnarray}
Using the expressions (\ref{eq:rhos}) and (\ref{zeta}), the Friedmann equation acquires the form
\begin{eqnarray}\label{HzSS}
H_D^2(z) &=&
H^2_0\,\left[\tilde{\Omega}^0_{M}\,\left(1+z\right)^{\alpha}+\tilde{\Omega}_{D}^{0}\,\zeta(z)\right]\,.
\end{eqnarray}
On the other hand we have the {\em variable CC picture}, which
describes the kind of models studied in this paper. Any of these
models incorporates the matter-radiation component, a variable
$\Lambda$ component and, possibly although not necessarily, a
variable gravitational coupling $G$. The variable $\Lambda$ model
represents a modification of Einstein equations of General
Relativity which maintains its geometrical interpretation. The
dynamical equations for gravity are given by
\begin{equation}
\label{eq:Einvarlam} R_{\mu \nu} -\frac{1}{2} g_{\mu \nu} R = 8\pi
G\,T_{\mu\nu}+\,g_{\mu\nu}\Lambda\equiv 8 \pi G (T_{\mu \nu} +
g_{\mu \nu} \rho_{\Lambda}) \, ,
\end{equation}
where $T_{\mu \nu}$ stands for the energy-momentum tensor of
matter. This equation demonstrates that the full covariance is
maintained even if $\Lambda$ and $G$ acquire space-time
variability. Here $\rho_{\Lambda}$ is the energy density
associated to $\Lambda$. In the framework of the FRW metric,
$\rho_{\Lambda}$ and $G$ depend on the cosmic time only, in
accordance with the cosmological principle. Friedmann's equation
is straightforwardly obtained from (\ref{eq:Einvarlam}) and reads
\begin{equation}\label{CCpicture}
H_{\Lambda}^2=\frac{8\pi G}{3}(\rho+\rho_{\Lambda})\,.
\end{equation}
The general Bianchi identity of the Einstein tensor leads to the
covariant conservation law
\begin{equation}
\label{eq:covBianchi} \nabla^{\mu} \left[G (T_{\mu \nu} + g_{\mu \nu} \rho_{\Lambda}) \right]=0 \, ,
\end{equation}
which for FRW metric acquires the form
\begin{equation}\label{BianchiGeneral}
\frac{d}{dt}\,\left[G(\rho+\rho_{\Lambda})\right]+3\,G\,H_{\Lambda}\,(\rho+p)=0\,.
\end{equation}
This ``mixed" conservation law connects the variation of
$\rho_{\Lambda}$, $G$ and $\rho$, where the scaling of the
matter-radiation density $\rho$ may be non-canonical in this
picture (in contrast to $\rho_s$ in the DE picture). In this
paper we consider a very broad class of models, just assuming the
variability of the aforementioned quantities, without specifying
the fundamental origin of such a variation. A number of variable
CC models of various kinds \cite{CCvariable1}, and the
renormalization group (RG) models of running $\rho_{\Lambda}$ and
$G$ \cite{JHEPCC1,Babic,Reuter} provide the basis  for the
variability of these cosmological parameters. For example, the RG
models  \cite{JHEPCC1,Babic} do not determine the time dependence
of $\rho_{\Lambda}$ and $G$ directly, but indirectly specifying
them in terms of other cosmic dynamic quantities (matter density
$\rho$, Hubble parameter $H$, etc):
\begin{equation}\label{variableCCG}
\rho_{\Lambda}(z)=\rho_{\Lambda}(\rho(z),H(z),...)\,,\ \ \ \ \ G(z)=G(\rho(z),H(z),...)\,.
\end{equation}
These functions usually have a monotonic dependence when
expressed as functions of cosmic time or redshift. The relations
(\ref{variableCCG}), with the general conservation law
(\ref{BianchiGeneral}), lead to the complete solution of the
variable CC cosmological model. Using this solution the general
expression for the Hubble parameter becomes
\begin{equation}\label{HLambda}
H_{\Lambda}^2(z)=H^2_0\,\left[\Omega_{M}^0\,f_{M}(z;r)(1+z)^{ \alpha}
+\Omega_{\Lambda}^0\,f_{\Lambda}(z;r)\right]\,.
\end{equation}
Here $f_{M}$ and $f_{\Lambda}$ are known functions of redshift
which may also depend on parameters $r=r_{1}, r_{2}, \dots$
originating from the fundamental dynamics. They generally have a
nontrivial dependence on $z$ and only in the case of $\Lambda$CDM
cosmology they satisfy $f_{M}=f_{\Lambda}=1$. Furthermore, they
also fulfill the conditions $f_{M}(0,r)=1$ and
$f_{\Lambda}(0,r)=1$ in accordance with the cosmic sum rule
$\Omega_{M}^{0}+\Omega_{\Lambda}^{0}=1$. Notice that in general
the two sets of cosmological parameters in the two pictures
(\ref{HLambda}) and (\ref{HzSS}) will be different, e.g.
$\Delta\Omega_{M}\equiv \Omega_{M}^0-\tilde{\Omega}_{M}^0\neq 0$,
because they correspond to two different fits of the same data.

\section{Matching of pictures and effective dark energy equation of state}

The two pictures presented in the preceding section may be
considered as two separate, general DE models. In the remainder of
the paper we, however, assume that they are equivalent
descriptions of the same cosmological evolution. More precisely,
we study the effective DE dynamics associated to the variable CC
model through the procedure named the matching of pictures. The
matching of pictures requires that the expansion history of the
universe is the same in both pictures, i.e. that their Hubble
functions are equal, $H_{D}=H_{\Lambda}$ at all times. In this
way, for a known dynamics  of the variable CC model, an effective
DE can be constructed. A number of general results for the
behavior of the effective DE density can be obtained with
interesting implications to the observational data. The
aforementioned matching of the two pictures gives the following
equality connecting the dynamical cosmological quantities in both
pictures, $G(\rho + \rho_{\Lambda}) = G_{0}(\rho_{s}+\rho_{D})$.
Using $H \, dt = -dz/(1+z)$, the general Bianchi identity
(\ref{BianchiGeneral}) can be cast in the form
\begin{equation}\label{droro}
{(1+z)}\,\,d(\rho_s+\rho_D)=\alpha\,\left(\rho_s+\rho_D-\xi_{\Lambda}\right)\,{dz}\,,
\end{equation}
where we have introduced
\begin{equation}\label{xiL}
\xi_{\Lambda}(z)=\frac{G(z)}{G_0}\,\rho_{\Lambda}(z)\,.
\end{equation}
Next we use the scaling law (\ref{eq:conservNR}) for $\rho_{s}$
and we arrive at the compact form for the redshift evolution of
the effective DE density in the variable CC picture:
\begin{equation}\label{drdz}
\frac{d\rho_D(z)}{dz}=\alpha\,\frac{\rho_D(z)-\xi_{\Lambda}(z)}{1+z}\,.
\end{equation}
The integration of (\ref{drdz}) readily yields a closed form
expression for the effective DE density
\begin{equation}\label{IF}
\rho_D(z)=\left(1+z\right)^{\alpha}\left[\rho_D(0)-\alpha
\int_0^z\frac{dz'\,\xi_{\Lambda}(z')}{(1+z')^{(\alpha+1)}}\right]\,.
\end{equation}
Expanding $\xi_{\Lambda}(z)$ around $z=-1$ one can see that
$\rho_D(z) \rightarrow \xi_{\Lambda}(z)$ at sufficiently late time
(i.e. when $z \rightarrow -1$). Finally, the effective EOS
parameter for the variable CC model easily follows from
(\ref{zeta}) and (\ref{drdz})
\begin{equation}\label{we2}
w_{\rm
eff}(z)=-1+\,\frac13\,(1+z)\,\frac{1}{\rho_D}\frac{d\rho_D}{dz}=
-1+\frac{\alpha}{3}\,\left(1-\frac{\xi_{\Lambda}(z)}{\rho_D(z)}\right)
\,,
\end{equation}
and it is seen to depend on the structure of the quantities
(\ref{xiL}) and (\ref{IF}). From this expression we immediately
see that in this kind of models $w_{\rm eff}$ can cross the
$w_{\rm eff} = -1$ line as soon as $\xi_{\Lambda}$ equals
$\rho_{D}$. This observation reinforces the role of the effective
DE density $\rho_{D}$ in the study of the CC boundary crossing in
the dark energy picture\,\footnote{For other recent theoretical
approaches to the $w_{\rm eff} = -1$ boundary crossing see e.g.
\cite{cross}.}.

\section{Effective quintessence and phantom behavior of cosmological models with variable $\Lambda$ and $G$}

For the following considerations it will be convenient to recast
the solution of the ODE (\ref{drdz}) in a different way.  Let
$z^{*}$ be the redshift value at which
$\xi_{\Lambda}(z^*)=\rho_D(z^*)$, i.e. $w_{\rm eff}(z^*)=-1$. Then
it is easy to show that (\ref{drdz}) is solved by the density
function
\begin{equation}\label{IF2}
\rho_D(z)=\xi_{\Lambda}(z)-\left(1+z\right)^{\alpha}\,
\int_{z^{*}}^z\frac{dz'}{(1+z')^{\alpha}}\frac{d\xi_{\Lambda}(z')}{dz'}\, .
\end{equation}
Quite remarkably, one can prove that a value $z^*$ {\em always}
exists near our present time; namely in the recent past, in the
immediate future or just at $z^*=0$. The proof of this assertion
is obtained by straightforward calculation using: i) the matching
condition $H_D=H_{\Lambda}$ and ii) the differential constraint
that the general Bianchi identity (\ref{BianchiGeneral}) imposes
on the functions $f_{\Lambda}$ and $f_{M}$ in (\ref{HLambda}). In
this way one obtains the following relation
\begin{equation}\label{dzeta}
\frac{d\rho_D(z)}{dz}=\frac{\alpha\,(1+z)^{\alpha-1}}{1-\tilde{
\Omega}_M^0}\rho_D(0)
\left(\Omega_M^0\,f_M(z;r)-\tilde{\Omega}_M^0\right)\,.
\end{equation}
From this the proof is obvious. Indeed, we have already mentioned
that at the present epoch function $f_M$ satisfies the condition
$f_M(0,r)=1$. Moreover, the parameter difference $\Delta \Omega_M
= \Omega_M^0 - \tilde{\Omega}_M^0$ should not be large because
the two pictures (DE and CC) describe the same physics. Hence from
the continuity of $f_M$ it is clear that there must be a point
$z^*$ close to 0 where (\ref{dzeta}) vanishes, that is to say a
point that satisfies $w_{\rm eff}(z^*)=-1$. The advantage of the
formulation (\ref{IF2}) becomes evident when one calculates the
slope of the $\rho_{D}$ function:
\begin{equation}\label{dIF2}
\frac{d\rho_D(z)}{dz}=-\alpha\,\left(1+z\right)^{\alpha-1}
\int_{z^{*}}^z\frac{dz'}{(1+z')^{\alpha}}\frac{d\xi_{\Lambda}(z')}{dz'}\,.
\end{equation}
This compact expression reveals some counterintuitive and general
aspects of the effective DE density evolution for variable CC
models in which $\xi_{\Lambda}(z)$ is a monotonous function of
$z$. Intuitively one would expect that for $\xi_{\Lambda}(z)$
growing/decreasing with $z$ (decreasing/growing with expansion)
$\rho_{D}$ should be quintessence-like/phantom-like. The
expression (\ref{dIF2}) shows that this is not the case. Namely,
for $\xi_{\Lambda}(z)$ growing with $z$, $\rho_{D}$ decreases with
$z$ for $z>z^*$, i.e. in this redshift interval has the
phantom-like characteristics. Only for $z<z^*$ $\rho_{D}$ behaves
as quintessence. Analogously, for $\xi_{\Lambda}(z)$ decreasing
with $z$, $\rho_{D}$ behaves like quintessence for $z>z^*$,
whereas only for $z<z^*$ it becomes phantom-like. Interestingly
enough, if $z^*$ lies near our recent past (see
Fig.\,\ref{plot}a), the last case could just reflect the
experimental indications of DE phantom-like behavior suggested by
recent analyses of the data\,\cite{transition}. A concrete
framework describing this last possibility is outlined in Section
\ref{RGmodel}. These results illustrate that in variable CC
models, the behavior of effective DE density is generally not
determined by the CC only, but by the joint behavior of all
quantities entering the general Bianchi identity
(\ref{BianchiGeneral}). Especially interesting results are
obtained when in the variable CC models the matter component
$\rho$ is separately conserved, i.e. when it also satisfies
(\ref{eq:conservNR}). In this case (\ref{BianchiGeneral}) implies
${d\xi_{\Lambda}}/{dt}=-(\rho/G_0)\,{dG}/{dt}$, which from
(\ref{dIF2}) results in
\begin{equation}\label{varG}
\frac{d\rho_D}{dz}=\alpha (1+z)^{\alpha-1} \,\frac{\rho(0)}{G_0}\ [G(z)-G(z^{*})]\,.
\end{equation}
Thus in this case the properties of $\rho_{D}$ depend only on the
scaling of $G$ with redshift, e.g. if $G$ is asymptotically free
and $z^{*}>z$, then $\rho_D$ behaves effectively as quintessence
(${d\rho_D}/{dz}>0$), whereas if $G$ is infrared-free then
$\rho_D$ behaves phantom-like (${d\rho_D}/{dz}<0$).

\section{Effective dark energy picture of the RG model}
\label{RGmodel}

\begin{figure}[t]
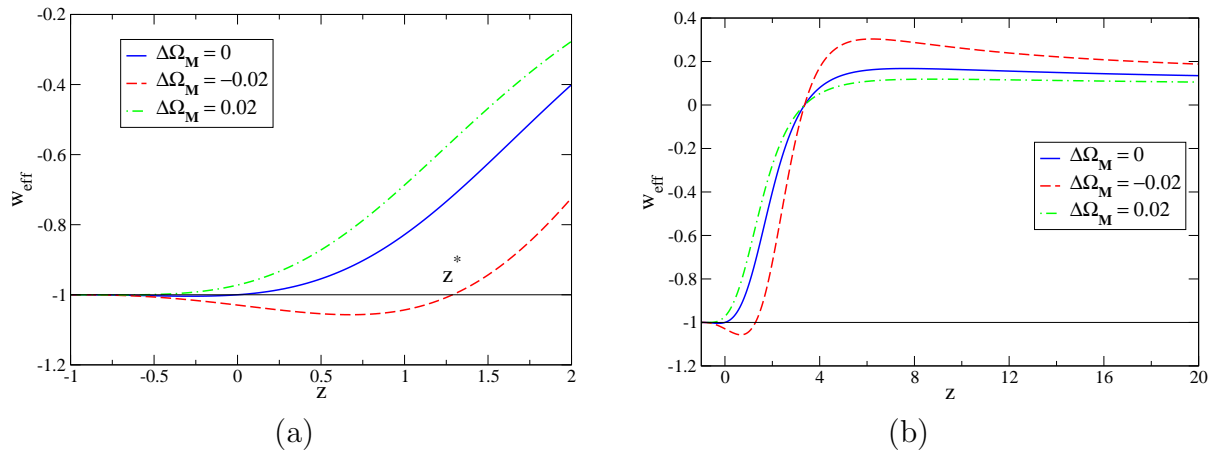

    \begin{tabular}{cc}
      \resizebox{0.48\textwidth}{!}{\includegraphics{minusnu2zmod.eps}} &
      \hspace{0.3cm}
      \resizebox{0.48\textwidth}{!}{\includegraphics{minusnu20z.eps}} \\
      (a) & (b)
    \end{tabular}
    \caption{\textbf{(a)}\ Numerical analysis of the effective EOS parameter
 $w_{\rm eff}$, Eq.\,(\protect\ref{wpflat1}), as a function of the redshift for
fixed $\nu=-\nu_0<0$, and for various values of $\Delta\Omega_M$.
The Universe is assumed to be spatially flat ($\Omega_K^0=0$)
with the standard parameter choice
$\Omega_M^0=0.3\,,\Omega_{\Lambda}^0=0.7$; \textbf{(b)} Extended
$z$ range of the plot (a). We see that for $\Delta\Omega_M<0$
there exists a transition point  $z^{*}$ near our recent past:
namely, the one corresponding to the crossing of the CC barrier
$w_{\rm eff}=-1$ by the lowest curve in the figures.}
  \label{plot}
\end{figure}

As a concrete illustration of the general procedure we have
developed for obtaining the effective DE properties, in this
section we consider the analysis \cite{SS1} of the
renormalization group cosmological model of \cite{RGTypeIa}
characterized by $G =\mathrm{const}$ and the scaling law
$\rho_{\Lambda}=C_1+C_2 H^2$ for the CC. Here
$C_1=\rho_{\Lambda,0}-(3 \nu H_0^2)/(8 \pi G)$ and $C_2=(3
\nu)/(8 \pi G)$, where $\nu$ is the single free parameter of the
model --a typical value is $|\nu|=\nu_0\equiv
1/12\pi$\,\cite{SS1}. This model is fully analytically tractable
and relatively simple expressions for $w_{\rm eff}$ can be
obtained. In this particular case it is clear that (\ref{xiL})
reads $\xi_{\Lambda}(z)=\rho_{\Lambda}(z)$. Therefore, for the
flat universe case the effective parameter of EOS obtained from
(\ref{IF}) and (\ref{we2}) is
\begin{equation}\label{wpflat1}
\hspace{-2cm} w_{\rm eff}(z)
=-1+(1-\nu)\,\frac{\Omega_M^0\,(1+z)^{3(1-\nu)}-\tilde{\Omega}_M^0\,(1+z)^3}
{\Omega_M^0\,[(1+z)^{3(1-\nu)}-1]-(1-\nu)\,[\tilde{\Omega}_M^0\,(1+z)^3-1]}\,.
\end{equation}
For $|\nu|\ll 1$ we may expand the previous result in first order
in $\nu$. Assuming $\Delta\Omega_{M}\equiv
\Omega_{M}^0-\tilde{\Omega}_{M}^0=0$ we find
\begin{equation}\label{expwq}
w_{\rm
eff}(z)\simeq-1-3\,\nu\frac{\Omega_M^0}{\Omega_{\Lambda}^0}\,(1+z)^3\,\ln(1+z)\,.
\end{equation}
This result reflects the essential qualitative features of the
general analysis presented in the previous sections. For $\nu>0$,
Eq.\,(\ref{expwq}) clearly shows that we can get an (effective)
phantom-like behavior ($w_{\rm eff}<-1$) and for $\nu<0$ we have
(effective) quintessence behavior. We see that this variable CC
model can give rise to two types of very different behaviors by
just changing the sign of a single parameter. However, one can
play with more parameters if desired. Indeed, as we have seen the
cosmological parameters in the two pictures (DE versus CC
picture) will generally be different ($\Delta\Omega_M\neq 0$).
Fig.\ref{plot} shows in a patent manner that in this case, even
for $\nu<0$, the variable CC model may exhibit phantom behavior
due to the existence of a transition point $z^{*}$ in our recent
past.
\vspace{0.0cm}
\section{Conclusions}
We have shown that a model with variable $\Lambda$ (and possibly
of $G$) generally leads to a non-trivial effective EOS, thus
mimicking a canonical dynamical DE model (i.e. one with conserved
DE density). That EOS can effectively appear as quintessence and
even as phantom energy. Moreover, we have proven that there
\textit{always} exists a transition point $z^{*}$ near $z=0$
where $w_{\rm eff}(z^{*})=-1$. If this point lies in our recent
past (as illustrated in Fig.\,\ref{plot}a) there could have been
a recent transition into an (effective) phantom regime $w_{\rm
eff}(z)<-1$, as suggested by several analysis of the data. If,
however, experiments would unambiguously indicate $w_{\rm
eff}(z)>-1$ we could equally well interpret this in our framework,
for $z^{*}$ could just be in our immediate future. We conclude
that variable $(\rho_{\Lambda},G)$ models may account for the
observed evolution of the DE, without need of invoking any
combination of fundamental quintessence and phantom fields. The
eventual determination of an empirical EOS for the DE in the next
generation of precision cosmology experiments should keep in mind
this possibility.

\textit{Acknowledgements}. This work has been supported in part
by MEC and FEDER under project 2004-04582-C02-01, and also by
DURSI Generalitat de Catalunya under project 2005SGR00564. The
work of HS is financed by the Secretaria de Estado de
Universidades e Investigaci\'on of the Ministerio de Educaci\'on
y Ciencia of Spain. HS thanks the Dep. ECM of the Univ. of
Barcelona for the hospitality.

\end{document}